\documentclass[aps,pra,twocolumn,showpacs,floatfix]{revtex4} 
\usepackage{amsmath,amssymb,isolatin1,graphicx,times}

\newcommand{\be}{\begin{equation}}
\newcommand{\ee}{\end{equation}}
\newcommand{\bea}{\begin{eqnarray}}
\newcommand{\eea}{\end{eqnarray}}
\newcommand{\bw}{\begin{widetext}}
\newcommand{\ew}{\end{widetext}}

\newcommand{\kommentar}[1]{}

\begin{document}
 
\title{Slow transport by continuous time quantum walks}
\author{Oliver M{\"u}lken}
\email{oliver.muelken@physik.uni-freiburg.de}
\author{Alexander Blumen}
\email{blumen@physik.uni-freiburg.de}
\affiliation{Theoretische Polymerphysik, Universit\"at Freiburg,
Hermann-Herder-Straße 3, 79104 Freiburg i.Br., Germany}

\date{\today} 
\begin{abstract}
Continuous time quantum walks (CTQW) do not necessarily perform better
than their classical counterparts, the continuous time random walks
(CTRW).  For one special graph, where a recent analysis showed that in a
particular direction of propagation the penetration of the graph is faster
by CTQWs than by CTRWs, we demonstrate that in another direction of
propagation the opposite is true; In this case a CTQW initially localized
at one site displays a slow transport. We furthermore show that when the
CTQW's initial condition is a totally symmetric superposition of states of
equivalent sites, the transport gets to be much more rapid.
\end{abstract}
\maketitle

The transfer of information over discrete structures (networks) which are
not necessarily regular lattices has become a topic of much interest in
recent years. The problem is relevant to many distinct fields, such as
polymer physics, solid state physics, biological physics and quantum
computation, see Refs.\
\cite{Bouchaud1990,albert2002,dorogovtsev2002,kempe2003,vankampen,weiss}
for reviews.  In particular, quantum mechanics seems to allow a much
faster transport than classically possible.  Thus, recent studies of
quantum walks on graphs show that these often outperform their classical
counterparts, i.e., in terms of the penetrability of the graph, for an
overview see Ref.\ \cite{kempe2003} and references therein. We recall that
the extension of classical random walks to the quantum domain is not
unique. There exist different variants of quantum walks, such as discrete
\cite{aharonov1993} and continuous time \cite{farhi1998} versions, which
are not equivalent to each other. Here we focus on walks in continuous
time.

Walks occur over graphs which are collections of connected nodes.  To each
graph corresponds a discrete Laplace operator (sometimes also called
adjacency or connectivity matrix), ${\bf A} = (A_{ij})$. Here the
non-diagonal elements $A_{ij}$ equal $-1$ if nodes $i$ and $j$ are
connected by a bond and $0$ otherwise. The diagonal elements $A_{ii}$
equal the number of bonds which exit from node $i$, i.e., $A_{ii}$ equals
the functionality $f_i$ of the node $i$. 

Classically, assuming the transmission rates of all bonds to be equal, say
$\gamma$, the continuous-time random walk (CTRW) is governed by the master
equation \cite{weiss}
\be
\frac{\rm d}{{\rm d} t} p_{j}(t) = \sum_k T_{jk} \ p_{k}(t),
\label{mast_eq0}
\ee
where $p_j(t)$ is the probability to find at time $t$ the walker at node 
$j$. ${\bf T} = (T_{jk})$ is the transfer matrix of the walk, which is
related to the adjacency matrix by ${\bf T} = - \gamma {\bf A}$.  Equation
(\ref{mast_eq0}) is spatially discrete, but it also admits extensions to
continuous spaces, e.g., leading to the disordered Lorentz gas model,
which describes the dynamics of an electron through a disordered
substrate~\cite{muelken2004}.

We stick with the spatially discrete situation and let now the CTRW start
from node $k$, i.e., we set $p_k(0) = \delta_{jk}$. Denoting by
$p_{jk}(t)$ the conditional probability of being at node $j$ at time $t$
when starting at node $k$ at $t=0$ leads to
\be
\frac{\rm d}{{\rm d} t} p_{jk}(t) = \sum_l T_{jl} \ p_{lk}(t),
\label{mast_eq}
\ee
which is another way to write Eq.(\ref{mast_eq0})
\cite{vankampen,weiss}. Given the linearity of these equations, their
solution involves a simple integration. For Eq.(\ref{mast_eq}) we have
formally
\be
p_{jk}(t) = \langle j | e^{{\bf T} t} | k \rangle.
\label{cl_prob}
\ee

\begin{figure}
\centerline{\includegraphics[width=0.9\columnwidth]{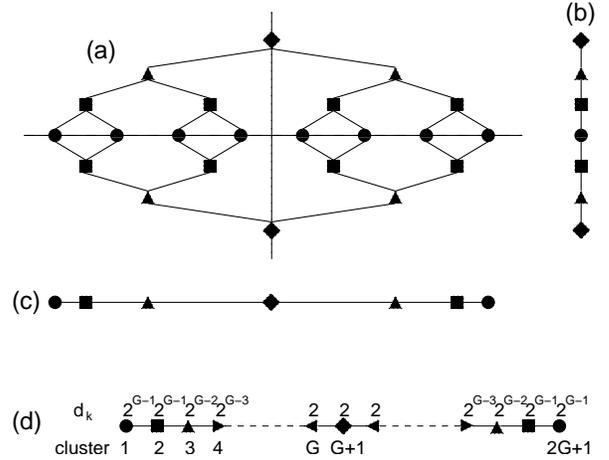}}
\caption{
(a) Graph consisting of two Cayley trees of generation $G=3$.  (b)
horizontal projection of the graph following Ref.\ \cite{childs2002}, (c)
vertical projection of the same graph. (d) Vertical projection of a
similar graph, obtained from two Cayley trees of general generation $G$,
indicating the new nodes (clusters) and the $d_k$, see text for details.}
\label{cayley_comb}
\end{figure}

We now turn to one quantum mechanical extension of the problem, the
so-called continuous-time quantum walk (CTQW). CTQWs are obtained by
assuming the Hamiltonian of the system to be ${\bf H} = - {\bf T}$
\cite{farhi1998,childs2002}. Then the basis vectors $|k\rangle$ associated
with the nodes $k$ of the graph span the whole accessible Hilbert space.
In this basis the Schr{\"o}dinger equation reads
\be
i \frac{\rm d}{{\rm d} t} | k \rangle = {\bf H} | k \rangle,
\label{sgl}
\ee
where we set $\hbar\equiv1$. The transition amplitude $\alpha_{jk}(t)$ from
state $| k \rangle$ at time $0$ to state $|j\rangle$ at time $t$ is then 
\be
\alpha_{jk}(t) = \langle j | e^{-i {\bf H} t} | k \rangle.
\label{qm_ampl}
\ee
According to Eq.(\ref{sgl}) the $\alpha_{jk}(t)$ obey
\be
i \frac{\rm d}{{\rm d} t} \alpha_{jk}(t) = \sum_l H_{jl}
\alpha_{lk}(t).
\label{sgl_ampl}
\ee
The inherent difference between Eq.(\ref{cl_prob}) and Eq.(\ref{qm_ampl})
is, apart for the imaginary unit, the fact that classically $\sum_j
p_{jk}(t) = 1$, whereas quantum mechanically $\sum_j |\alpha_{jk}(t)|^2
=1$ holds.

We turn now to the graph displayed in Fig.\ \ref{cayley_comb}(a). The
graph is obtained from two finite Cayley trees of generation $G$ which
have a common set of end nodes along the horizontal symmetry axis
indicated in Fig.\ \ref{cayley_comb}(a), \cite{farhi1998,childs2002}. For
the nodes on the axis as well as for the top and bottom nodes the
connectivity is $f=2$, whereas for all other nodes $f=3$. 

The authors of Refs.\ \cite{farhi1998} and \cite{childs2002} have analyzed
CTQWs over the graph given in Fig.\ \ref{cayley_comb}(a), focussing on
walks which start at the top node, and looking for the amplitude,
Eq.(\ref{qm_ampl}), of being at the bottom node at time $t$. The problem
can then be simplified by considering only states which are totally
symmetric superpositions of states $|k\rangle$ involving all the nodes $k$
in each row of Fig.~\ref{cayley_comb}(a), as indicated schematically in
Fig.~\ref{cayley_comb}(b). The transport gets then mapped onto a
one-dimensional CTQW \cite{childs2002}.

\begin{figure}
\centerline{\includegraphics[width=0.9\columnwidth]{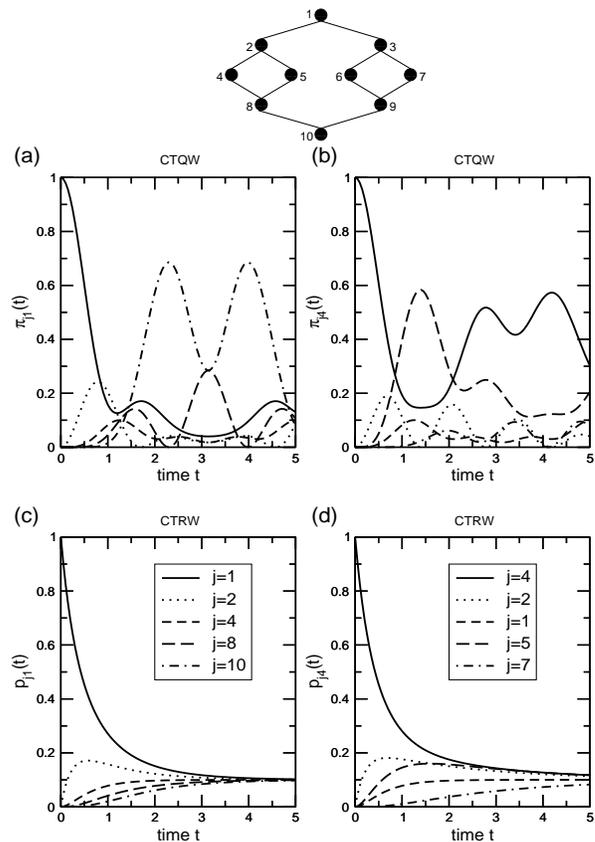}}
\caption{
Top: Graph obtained from two Cayley-trees of generation $G=2$. Below:
Probabilities for the CTQW, (a) and (b), and for the CTRW, (c) and (d),
to be at node $j$ after time $t$ when starting at time $0$ from node 1 or
from node 4. Left
column, (a) and (c): Starting node is the top node 1. Right column, (b)
and (d): Starting node is the leftmost node 4. The time is given in units
of the inverse transmission rate $\gamma^{-1}$, see text following Eq.(\ref{mast_eq0}).}
\label{cayley_10}
\end{figure}

Given that CTQWs obey time inversion, so that they never reach a limiting
distribution, one uses the quantity \cite{aharonov2001}
\be
\chi_{jk}  = \lim_{T\to\infty} \frac{1}{T} \int_0^T dt \ |\alpha_{jk}(t)|^2.
\label{limit_distr}
\ee
to compare the efficiency of CTQWs to that of CTRWs. We will show in the
following that the $\chi_{jk}$ may depend strongly on the initial state.
Now, as shown in \cite{childs2002}, based on
Eq.(\ref{limit_distr}), the CTQW's probability of being at the bottom node
when starting at the top node is considerably larger than that of CTRWs.

One legitimate question to ask now is: What happens if one considers on
the same graph CTQWs which start at the leftmost node and end at the
rightmost node? As we proceed to show, it turns out that then the
transport by CTQWs gets to be much slower than the transport by CTRWs. We
start by focussing on the full solution of Eq.(\ref{sgl_ampl}), for which
all the eigenvalues {\sl and} all the eigenvectors of ${\bf T}=-{\bf H}$
(or, equivalently, of ${\bf A}$) are needed.  For, say, the 22 nodes of
Fig.\ \ref{cayley_comb}(a) we have to solve the eigenvalue problem for
${\bf A}$ (or ${\bf T}$), which is a real and symmetric $22\times22$
matrix.  This is a well-known problem, also of much interest in polymer
physics \cite{blumen2003,blumen2004}, and many of the results obtained
there can be used for our problem here. 

We recall first that the matrix ${\bf A}$ is non-negative definite. Then,
for a structure like the one in Fig.~\ref{cayley_comb}(a), ${\bf A}$ has
exactly one vanishing eigenvalue, $\lambda_0=0$, the remaining eigenvalues
being positive.  Let $\lambda_n$ denote the $n$th eigenvalue of ${\bf A}$
and ${\bf \Lambda}$ the corresponding eigenvalue matrix. Furthermore, let
${\bf Q}$ denote the matrix constructed from the orthonormalized
eigenvectors of ${\bf A}$, so that ${\bf A} = {\bf Q}{\bf\Lambda}{\bf
Q}^{-1}$. Now the classical probability is given by
\be
p_{jk}(t) = \langle j| {\bf Q} e^{-t \gamma{\bf \Lambda}} {\bf Q}^{-1} | k
\rangle.
\label{cl_prob_full}
\ee
For the quantum mechanical transition probability it follows that
\be
\pi_{jk}(t) \equiv |\alpha_{jk}(t)|^2 = |\langle j| {\bf Q} e^{- i t \gamma {\bf
\Lambda}} {\bf Q}^{-1} | k \rangle|^2.
\label{qm_prob_full}
\ee

\begin{figure}
\centerline{\includegraphics[width=0.85\columnwidth]{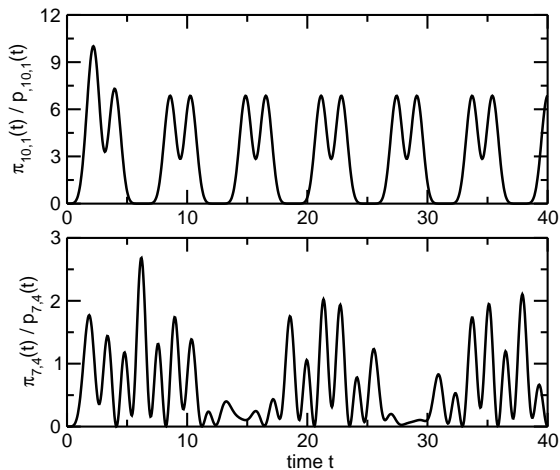}}
\caption{Ratios $\pi_{jk}(t)/p_{jk}(t)$ for
different directions of propagation, (a) top-bottom walk and (b)
left-right walk over time $t$, see Fig.\ref{cayley_10} for units and
details.}
\label{cayley_10_long}
\end{figure}

In order to determine numerically the corresponding eigenvalues and
eigenvectors of the matrix ${\bf A}$ for different graphs we have used the
standard software package MAPLE 7. We start by considering the smaller
graph, $G=2$, given at the top of Fig.\ \ref{cayley_10}. The figures show
the transition probabilities for CTQWs and CTRWs starting at the top node
1 (left column), which corresponds to the situation described in
\cite{childs2002}, or at the leftmost node 4 (right column). 
Remarkably, CTQWs starting at the top node reach the opposite node 10 very
quickly, see Fig.\ \ref{cayley_10}(a), much quicker than expected from the
CTRW behavior, Fig.\ \ref{cayley_10}(c).  However, for walks starting at
the leftmost node 4 and going to the rightmost node 7, the probabilities
for the CTQWs, Fig.\ \ref{cayley_10}(b), and the CTRWs, Fig.\
\ref{cayley_10}(d), get to be comparable. Furthermore, the CTQWs'
probabilities, $\pi_{4,4}(t)$ and $\pi_{1,1}(t)$, of return to the
starting node within the time interval depicted in Fig.~\ref{cayley_10}
are much higher if the walks start at the leftmost node 4 of the graph
instead of at the top node 1. On the other hand, for CTRWs there is not
much difference between starting at the leftmost or at the top node, only
that in the first case it just takes a little bit longer to reach an
uniform distribution, compare Fig.\ \ref{cayley_10}(c) and Fig.\
\ref{cayley_10}(d).

We now extend the time interval to $t=40$ and compare the efficiency of
the CTQW transport to the CTRW one.  In Fig.~\ref{cayley_10_long} we plot
for the top-bottom and the left-right walks the ratio of the quantum
mechanical probabilities $\pi_{jk}(t)$ to the classical ones $p_{jk}(t)$.
For top-bottom transport, depicted in Fig.\ \ref{cayley_10_long}(a), the
plot turns out to be highly regular, reflecting the high symmetry of the
underlying graph in the vertical direction. For left-right transport the
plot is less regular. Note the different scaling of the ordinates in the
two parts of Fig.~\ref{cayley_10_long}, which again stresses the
preferential role played by the transport in top-bottom direction.

In order to discuss what happens at even longer times, we proceed to
evaluate for the CTQW the limiting distributions $\chi_{jk}$ given by Eq.
(\ref{limit_distr}). For CTRWs the limit is simple: All $p_{jk}(t)$ tend
to the same constant, which is the inverse of the total number of nodes in
the graph, no matter where the CTRWs start.  For the CTQWs, however, this
is not the case, as can already be inferred from Figs.~\ref{cayley_10} and
\ref{cayley_10_long}. Hence, we compute $\chi_{jk}$ for the top-bottom and
for the left-right cases separately. Therefore, for the $G=2$ graph we
compute the eigenvalues and eigenvectors of the respective $10\times 10$
matrix ${\bf A}$ and the $\chi_{jk}$ by using MAPLE. Having the
appropriate eigenvalue matrix ${\bf\Lambda}$ and the matrix ${\bf Q}$
constructed from the orthonormalized eigenvectors we find with Eqs.\
(\ref{limit_distr}) and (\ref{qm_prob_full}) and the LinearAlgebra package of
MAPLE that for the top-bottom case
\be
\chi_{10,1} = 0.2644 > \frac{1}{10}, 
\label{limit_distr_10_cl}
\ee
whereas for the left-right case 
\be
\chi_{7,4} = 0.0545 <
\frac{1}{10}. 
\label{limit_distr_10_qm}
\ee
Thus, the limiting CTRW-probability, $1/10$, lies between $\chi_{7,4}$ and
$\chi_{10,1}$. The top-bottom CTQW is more, the left-right CTQW less
efficient than the corresponding CTRW.

\begin{figure}
\centerline{\includegraphics[width=0.85\columnwidth]{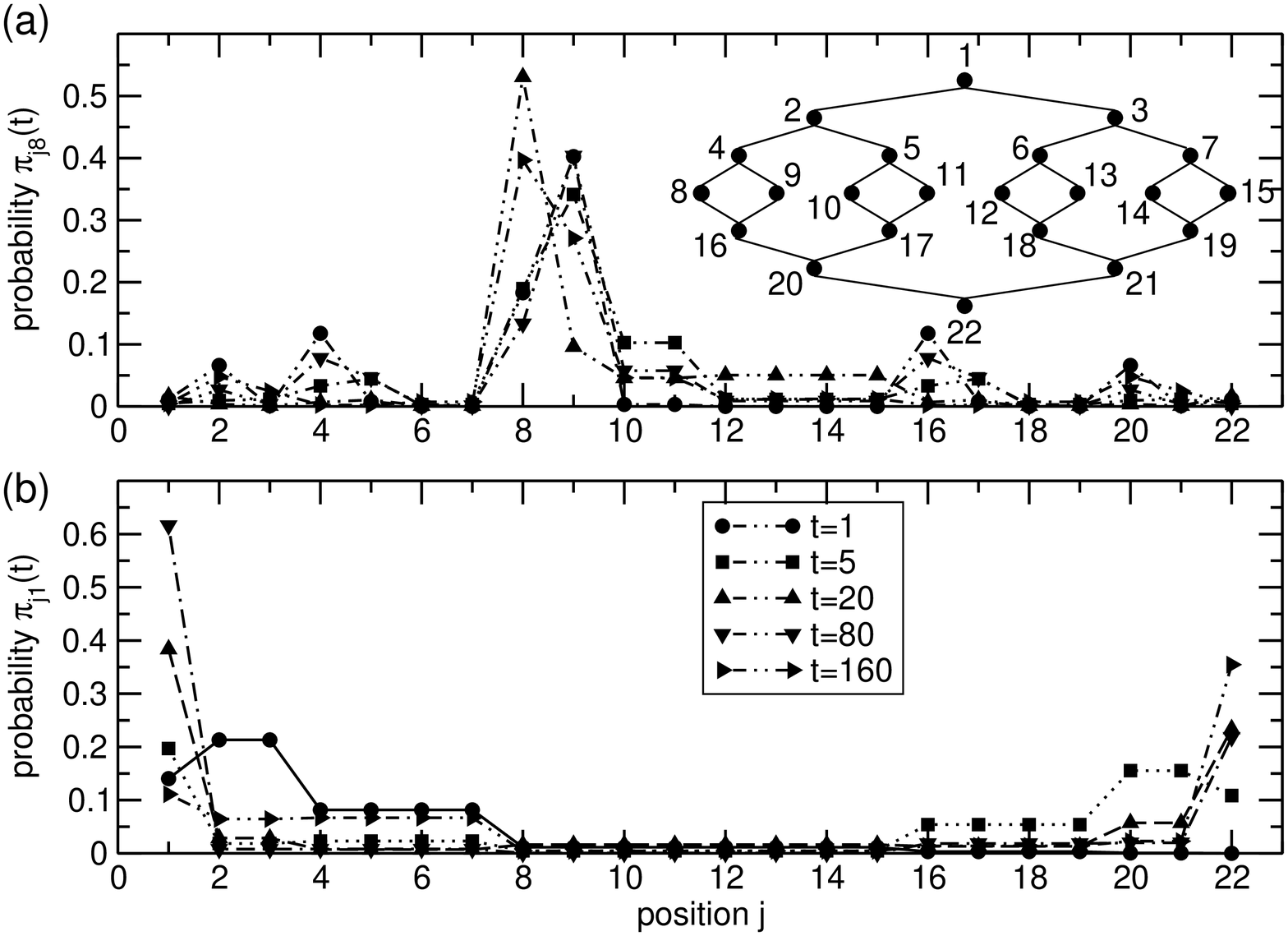}}
\caption{
Transition probabilities $\pi_{jk}(t)$ for a CTQW on the $G=3$ graph
starting at (a) the leftmost node, $\pi_{j8}(t)$, and (b) starting at the
top node, $\pi_{j1}(t)$, for times $t=1$, $5$, $20$, $80$, and $160$, see
Fig.\ref{cayley_10} for units and 
details.}
\label{cayley_22_pqm_time}
\end{figure}

\begin{figure}
\centerline{\includegraphics[width=0.85\columnwidth]{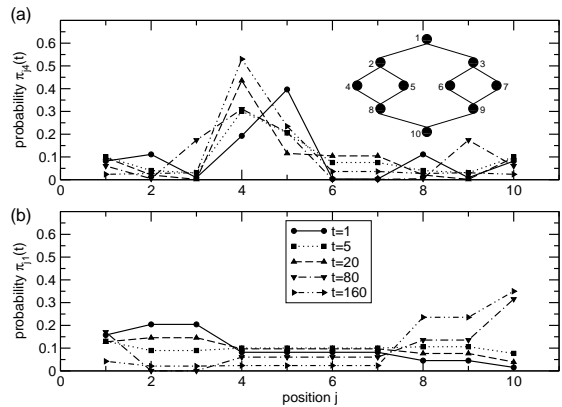}}
\caption{
Transition probability for a CTQW on the $G=2$ graph starting at (a) the
leftmost node, $\pi_{j4}(t)$, and (b) at the top node, $\pi_{j1}(t)$, for
times $t=1$, $5$, $20$, $80$, and $160$, see
Fig.\ref{cayley_10} for units and
details.}
\label{cayley_10_pqm_time}
\end{figure}

\begin{figure}
\centerline{\includegraphics[width=0.85\columnwidth]{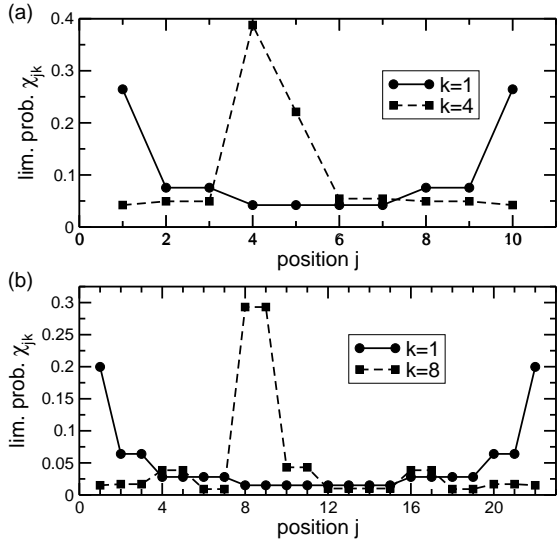}}
\caption{
Limiting probability $\chi_{jk}$ for a CTQW on the (a) $G=2$ graph and on
the (b) $G=3$ graph. Starting points are the top node, $k=1$ and the
leftmost nodes $f=4$ and $k=8$, respectively.}
\label{cayley_10_22_pqm_lim}
\end{figure}

In order to better visualize that the top-bottom and left-right $\chi_{jk}$
are very different, we show in Figs.\ \ref{cayley_22_pqm_time} and
\ref{cayley_10_pqm_time} the quantum mechanical transition probabilities
$\pi_{jk}(t)$ for all nodes $j$ when starting (a) at the leftmost node 8
and (b) at the top node 1; in these figures the time is displayed
parametrically.  Figure  \ref{cayley_22_pqm_time} is for the $G=3$ graph
and Fig.\ \ref{cayley_10_pqm_time} for the $G=2$ graph.  Now, even for the
small graphs considered here, we find differences in the transition
probabilities, which clearly depend on the initial node. For the $G=3$
graph consisting of $22$ nodes, the CTQW starting at the top node 1
spreads out rapidly over the whole graph. After a very short time
interval, there is a large probability to find the walker at the bottom
node 22, see Fig.\ \ref{cayley_22_pqm_time}(b). However, for the CTQW
starting at the leftmost node 8, we have up to times $t=160$ a high
probability of finding it in the left half of the graph, see Fig.\
\ref{cayley_22_pqm_time}(a). Therefore, the propagation of the CTQW is
strongly dependent on the starting node. For the smaller $G=2$ graph of
Fig.\ \ref{cayley_10_pqm_time}, which consists of $10$ nodes, the effect
is similar, but slightly less pronounced.

We illustrate the situation at very long times in Fig.\
\ref{cayley_10_22_pqm_lim}, where we display the limiting probabilities
$\chi_{jk}$ for the $G=2$ and the $G=3$ graphs, see
Eq.(\ref{limit_distr}).  For a CTQW starting at the top node 1 the
limiting probability distribution has its maximum at the end nodes of the
graphs, i.e. at nodes 1 and 10 for $G=2$, and at nodes 1 and 22 for $G=3$.
For a CTQW starting at the leftmost node, $k=4$ for $G=2$ and $k=8$ for
$G=3$, the limiting probability distribution shows a strong maximum around
the starting node.

Other initial conditions for the CTQW are, indeed, possible, especially
when considering the high symmetry of the underlying graphs.  Note that,
using for instance the site enumeration of Fig.\ \ref{cayley_22_pqm_time},
a CTQW from node $8$ to node $15$ is equivalent to a CTQW from, say, node
$10$ to node $14$.  The graph's symmetry suggests to collect groups of
such nodes into clusters, while focussing on the transport from left to
right. It is then natural to view the nodes 8, 9, 10, and 11 as belonging
to the first cluster.  The second cluster consists then of the nodes 4, 5,
16, and 17, all of which are directly connected by one bond to the nodes
of the first cluster. The nodes 2 and 20 of the third cluster are all
nodes directly connected by one bond to the nodes of the second cluster,
while at the same time not belonging to the first cluster. In general, all
the nodes of the $(k+1)$st cluster are connected by one bond to nodes of
the $k$th cluster and at the same time do not belong to the $(k-1)$st
cluster.

Let us denote the number of nodes in cluster $k$ by $d_k$. The transport
occurs now from a cluster to the next, by which the original graph gets
mapped onto a line in which one {\sl new} node corresponds to a {\sl
group} of original nodes of the graph. For a new node at position $k \in
[2,G]$ we find that $d_k = 2^{G-k+1}$, the same being true for the mirror
node value, i.e., $d_k=d_{2G+2-k}$.  Note that for the end nodes $d_1 =
d_{2G+1} = 2^{G-1}$, the same holds for the nodes next to them.  Moreover,
for the middle node $d_{G+1} = 2$. 

We now focus on the transport via the states which are totally symmetric,
normalized, linear state-combinations for all the original nodes in each
cluster. Thus, for the $k$th cluster, whose sites we denote by $n$, we
have as a new state
\be
| a_k \rangle = \frac{1}{\sqrt{d_k}}\sum_{n\in k} | n \rangle.
\label{line_states}
\ee

\begin{figure}
\centerline{\includegraphics[width=0.85\columnwidth]{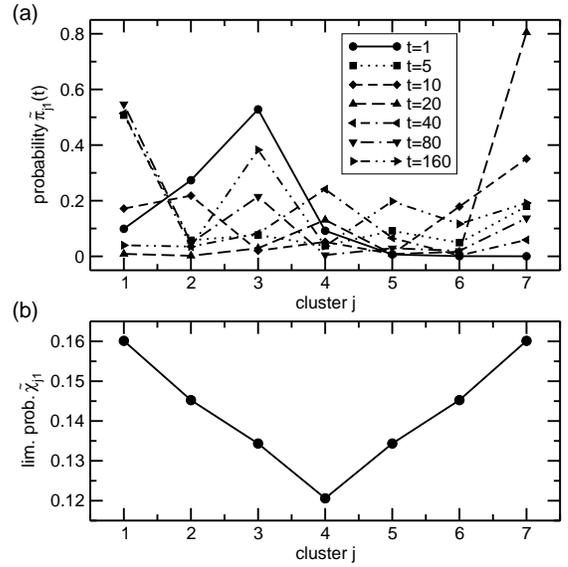}}
\caption{
(a) Transition probability $\tilde\pi_{j1}(t)$ for a CTQW between
different clusters $j$ of the $G=3$ graph. The CTQW  starts at the first
cluster, presented is the situation at
times $t=1$, $5$, $10$, $20$, $40$, $80$, and $160$, see
Fig.\ref{cayley_10} for units and
details.
(b) Limiting
probability $\tilde\chi_{j1}$ for a CTQW starting at the first cluster.}
\label{cayley_22_map_time}
\end{figure}

The CTQW is now determined by the new Hamiltonian $\tilde{\bf
H} = \gamma \tilde{\bf A}$, where the matrix elements of $\tilde{\bf A}$
are obtained from the new basis states $| a_k\rangle$ and from the matrix
${\bf A}$ through
\be
\tilde A_{jk} = \langle a_j | {\bf A} | a_k \rangle.
\ee
Given the properties of ${\bf A}$ and the construction of the
$|a_k\rangle$, Eq.(\ref{line_states}), $\tilde{\bf A}$ is a real and
symmetrical tridiagonal matrix, which implies a CTQW on a line. The
diagonal elements of $\tilde{\bf A}$ are given by
\be
\tilde A_{kk} = \langle a_k | {\bf A} | a_k \rangle = \frac{1}{d_k}
\sum_{{n \in k}\atop{n'\in k}} \langle n' | {\bf A} | n \rangle = f_n
\equiv f_k,
\ee 
where $f_k$ is the functionality of every node in the $k$th cluster.
For the sub- and super-diagonal elements of $\tilde{\bf A}$ we find
\bea
\tilde A_{k,k+1} &=& \tilde A_{k+1,k} = \langle a_k | {\bf A} | a_{k+1}
\rangle \nonumber \\
&=& \frac{1}{\sqrt{d_k d_{k+1}}} \sum_{{n \in k}\atop{n'\in k+1}}
\langle n' | {\bf A} | n \rangle = - \frac{b_k}{\sqrt{d_k d_{k+1}}},
\eea
where $b_k$ is the number of bonds between the clusters $k$ and $k+1$.
Now, except for the ends and the center of the graph, $b_k$ equals the
maximum of the pair $(d_k,d_{k+1})$. Between the central node
($d_{G+1}=2$) and its neighbors ($d_G=d_{G+2}=2$) the number of bonds is
$b_G = b_{G+2} = 2$. The number of bonds between the end node and its
neighbor is $b_1 = b_{2G+1} = 2 d_1 = 2^G$.

For the graph consisting of 22 {\sl original} nodes the new matrix
$\tilde{\bf A}$ is a tridiagonal $7\times7$ matrix, which can be readily
diagonalized.  The advantage of the procedure is clear: the new matrix
$\tilde{\bf A}$ depends on the number of clusters and grows with ($2G+1$),
whereas the full adjacency matrix, ${\bf A}$, grows with the total number
of nodes in the graph, namely with ($3\cdot2^G-2$). 

From Eq.(\ref{line_states}) the transition amplitude between the state
$|a_k\rangle$ at time 0 and the state $|a_j\rangle$ at time $t$ is given
by
\be
\tilde \alpha_{jk}(t) = \langle a_j | e^{- i \tilde{\bf H} t} | a_k \rangle 
= \langle a_j | \tilde{\bf Q} e^{-i\gamma\tilde{\bf \Lambda} t} \tilde{\bf
Q}^{-1}| a_k \rangle,
\ee
where $\tilde{\bf \Lambda}$ is the eigenvalue matrix and $\tilde{\bf Q}$
the matrix constructed from the orthonormalized eigenvectors of the new
matrix $\tilde{\bf A}$.

Now the quantum mechanical transition probabilities are given by
$\tilde\pi_{jk}(t) = |\tilde \alpha_{jk}(t)|^2$.  Figure
\ref{cayley_22_map_time}(a) shows the transition probabilities for CTQWs
over clusters. Remarkably now, and similar to Fig.\
\ref{cayley_22_pqm_time}(b), already in rather short periods of time such
CTQWs move from one end cluster to the other. The limiting probability
distribution, $\tilde\chi_{jk}$, which is depicted in Fig.\
\ref{cayley_22_map_time}(b), also supports this finding. Note that Fig.\
\ref{cayley_22_map_time}(b) again reflects the symmetry of the original
graph.

In conclusion, we have shown that CTQWs do not necessarily perform better
than their CTRWs counterparts. By focussing on a particular graph, we have
shown that the penetration of such a graph by CTQWs can be better or worse
than the one by CTRWs, depending on the initial state and on the
propagation direction under scrutiny.

Support from the Deutsche Forschungsgemeinschaft (DFG) and the Fonds der
Chemischen Industrie is gratefully acknowledged.

\end{document}